\documentclass[useAMS,usenatbib]{pasa}

\title[Substructure in Dwarf Galaxies]{Substructure in Dwarf Galaxies}
\author[Matthew Coleman]{Matthew Coleman,$^1$}
\affil{
$^1$Research School of Astronomy \& Astrophysics\\
Institute of Advanced Studies, The Australian National University\\
Cotter Road, Weston Creek, ACT 2611, Australia\\
coleman@mso.anu.edu.au}
\begin{document}

\maketitle

\label{firstpage}

\begin{abstract}

Recent years have seen a series of large-scale photometric surveys with the aim of detecting substructure in nearby dwarf galaxies.  Some of these objects display a varying distribution of each stellar population, reflecting their star formation histories.  Also, dwarf galaxies are dominated by dark matter, therefore luminous substructure may represent a perturbation in the underlying dark material.  Substructure can also be the effect of tidal interaction, such as the disruption of the Sagittarius dSph by the Milky Way.  Therefore, substructure in dwarf galaxies manifests the stellar, structural and kinematic evolution of these objects.

\end{abstract}


\section{Introduction}
The Galaxy is surrounded by a family of satellite systems, near enough to be resolved into individual stars and such that we are able to examine the internal structure and dynamics in each system.  These include the ten known dwarf spheroidal (dSph) galaxies as well as the two luminous irregular systems, the Large and Small Magellanic Clouds.  Historically, dSph galaxies were considered to be large, diffuse globular clusters; gas-free systems with a single-age stellar population dating back to the opening epoch of the universe.  Since the seminal study of the Carina dSph by \citet{mou83}, it has been found that the stellar populations of these systems range from mainly old (for example, Ursa Minor) to the mainly young (Leo I).  The colour-magnitude diagram can be used to isolate different stellar populations of a dSph galaxy in the search for substructure.  For example, the distribution of young stars within the Phoenix dIrr/dSph galaxy indicates recent star formation has propagated from east to west across the face of the dwarf \citep{spick99}.  That is, separating stellar populations can be used to determine the spatial star formation history within dwarf galaxies.

Dwarf galaxies are dominated by dark matter (DM); kinematic measurements indicate the visible component of a dwarf galaxy sits in the centre of a much larger dark halo.  \citet{aar83} measured the velocity dispersion of Draco to be $6.5$ km s$^{-1}$, resulting in a mass-to-light ratio measurement an order of magnitude above those for globular clusters.  It has since been found that all Galactic dSphs are highly massive given their luminosity, with central mass-to-light ratios of up to $\sim$$100$.  Dwarf galaxies are the darkest objects in the universe we can directly observe.  If we assume that stars act as a tracer of the underlying dark material, then the structure and kinematics of the luminous matter in a dwarf galaxy is driven by the dark halo.  Therefore, substructure in the stellar population may reflect corresponding substructure in the dark halo.

\section{Dwarf Galaxies in a CDM Universe}

The importance of dwarf galaxies in understanding DM and the formation of large-scale structure has become apparent in recent years.  Cold Dark Matter (CDM) simulations predict that small-scale clumps of dark matter ($\sim$10$^7$ M${}_{\odot}$) formed from fluctuations in the early universe.  These clumps merged to form dwarf-sized systems, which then accreted and merged to produce the large-scale structure we see today.  This implies two important points about the current crop of Galactic dSphs: (i) they are survivors from a previous epoch, and; (ii) they formed from accretion of smaller systems in the early universe.  The dSph galaxies are old systems with crossing times ($10^7 - 10^8$ Gyr) significantly less than their age, hence any substructure remaining from the initial formation of the dark halo is expected to have already dispersed.

Although Cold Dark Matter (CDM) simulations successfully reproduce the overall filamentary structure of the universe, they show discrepancies with observations at galactic scales.  \citet{tos03} gave a review on this subject.  One of the more concerning points is overprediction of dark halo numbers by at least an order of magnitude compared to observations of nearby systems \citep{moo99}.  It has been pointed out (and was re-iterated at this meeting) that the measurement of a dwarf galaxy's circular velocity does not correspond to the circular velocity measured from a simulated sub-halo of the Galaxy.  The circular velocity of simulated sub-halos plotted in Fig.\ 2 of \citet{moo99} was calculated as $v_c = \sqrt{Gm_b/r_b}$ where $m_b$ is the bound mass (visible and dark) within the bound radius $r_b$ of the substructure.  However, the circular velocities observed in dwarf galaxies contain only the mass within the {\it luminous} limiting radius, while the dark halo extends significantly beyond this boundary.  Also, star formation in smaller satellites is expected to be an inefficient process, such that the many smaller satellites predicted by CDM may not be easily visible.  While this is true, the ``missing satellite'' problem is still to be resolved.

\section{Sagittarius}
The dwarf galaxy in Sagittarius provides a unique opportunity to examine galaxy-galaxy interactions, as it represents the most extreme form of substructure in a dwarf galaxy: hierarchical merging.  Sagittarius is currently being torn apart by the gravitational influence of our own Galaxy.  Although it was discovered only a decade ago \citep{iba94} there has been a heavy concentration of effort to untangle the dynamical history of Sagittarius.  The most recent comprehensive review of such work can be found in \citet{maj03}.  By colour-selecting M giant stars from the 2MASS catalogue, \citet{maj03} have been able to trace the southern Sagittarius tidal tail over at least $150^{\circ}$ across the sky.  They also find evidence for the northern arm extending through the Galactic plane towards the North Galactic Pole, where it turns over and possibly crosses the Galactic plane again near the Sun.  The Sagittarius tidal tails are extensive, and appear to complete at least one loop in a polar orbit around the Galaxy.  See Figs.\ 3 and 4 of \citet{maj03} for a map of the Sagittarius tails.

This event is important towards our understanding of galactic dynamics.  Firstly, the tidal tails are very sensitive not only to the mass of the Galaxy, but also to the {\it distribution} of this mass.  Therefore, the shape and kinematics of the tidal tails can be used to measure the shape of the Galaxy's dark halo.  Secondly, some simulations predict that the dark halo of a satellite will help maintain its structure during a merging event \citep{may01,hel01}, however this process is not well understood.  Therefore, the mass loss rate of Sagittarius can be used to determine the amount of DM it contains, and also to find how efficiently a dark halo will conserve the luminous mass within a bound system while it is being perturbed.  \citet{maj03} predict a fractional mass loss rate for Sagittarius of $\sim$$15\%$ per gigayear, with at least $15\%$ of the mass of Sagittarius currently located in the tidal tails.  The recently discovered Canis Major dwarf remnant \citep{martin04} may represent the final stage of this merger process, possibly creating the old open clusters located in the outer region of the Galactic disk \citep{frinchaboy04}.

Five globular clusters are known to be associated with Sagittarius.  In the trailing tail, these include Terzan 7, Terzan 8, Arp 2, and most recently, Palomar 12 \citep{dinescu00}.  At the centre of Sagittarius lies M54, the second brightest of the Galactic globular clusters.  The globular clusters occupy the same phase-space region as the dSph, and are hence assumed to form part of the Sagittarius system.  Also, Sagittarius has a similar integrated magnitude to Fornax, which has five globular clusters \citep{hod61}.  Therefore, it seems plausible that the globular clusters were companions to the Sagittarius progenitor and have been tidally captured by the Milky Way.  The cluster M54, located at the centre of Sagittarius, may provide an important clue to the formation of nucleated dwarf galaxies.  Models predict the internal structure and velocity dispersion of a disrupted satellite should remain untouched by the gravitational potential of the host galaxy until the very end of the interaction (for example, \citealt{hel01}).  However, M54 may represent the nucleation of Sagittarius catalysed by the extreme tidal forces of the Galaxy, a possibility yet to be fully explored.  Conversely, the paucity of metals in M54 compared to the mean Sagittarius population \citep{dac95} may indicate the two objects are distinct bodies.

\section{Ursa Minor}
The structure of Ursa Minor dSph (UMi) is significantly distorted.  It is asymmetric about its core \citep{palma03}, and has a secondary density peak located just outside the core radius.  The stars which form this secondary density peak are not distinguished in colour and magnitude from the remainder of the UMi population \citep{kle98}.  Also, the radial velocity analysis by \citet{kwg03} found the sub-population is kinematically colder than the dSph, which implies the clump is long-lived.  \citet{kwg03} conclude that this stellar association represents an underlying DM clump which has survived the original merger formation of UMi, rather than distortion due to the Galactic tidal field.

However, the presence of DM in UMi is still under investigation.  The radial velocity studies of \citet{har94} and \citet{arm95} imply the dSph is dominated by DM, with a mass-to-light ratio in the range $30-80$, possibly the highest in the Galactic dSph population.  In contrast, the wide-field photometric surveys of \citet{spick01} and \citet{palma03} found UMi stars beyond the nominal tidal radius, indicating the system is tidally interacting with the Milky Way.  Thus, UMi does not appear to be in virial equilibrium, and the measured velocity dispersion may be artificially inflated.  \citet{spick03} have recalculated the mass-to-light ratio of UMi to be $\sim$12 using a method independent of the internal kinematics, and have also found that the tidal extension observed in this system cannot be produced under the presence of DM.

More recently, the kinematic study by \citet{wilkinson04} found the velocity dispersion drops at large radii, indicating the outer regions of UMi contains a kinematically cold population.  This argues against the interaction scenario, and suggests that UMi is an important test object for the DM-dominated formation scenarios for dwarf galaxies.

\section{Fornax}
Fornax appears to have unusually complex star formation and dynamical histories when considering its large distance from the centre of the Milky Way ($138$ kpc; \citealt{mat98}) and relatively gentle orbital parameters (Piatek 2003, personal communication).  Deep photometric studies by two groups \citep{ste98,sav00} have found that star formation in Fornax began approximately 12 Gyr ago and has continued intermittently almost to the present day; some stars are no older than 200 Myr.  Fornax does not appear to have any associated HI, hence it is unknown how the galaxy has managed to maintain an almost unbroken cycle of star formation.

The distribution of Fornax stars also reflect a complex dynamical history.  Previous studies have revealed possible excess Fornax stars beyond the tidal radius \citep{ih95}, and asymmetries in the distribution of stars in the inner regions.  The distribution of young stars (ages $< 2$ Gyr) is complex and appears to be concentrated towards the centre of Fornax \citep{ste98}.  First results from a wide-field medium-depth survey indicate Fornax substructure may be particularly complex compared to other Galactic dSph systems.  \citet{col04} detected an association of stars with ages $\sim$2 Gyr near the core radius of Fornax which appear to be separated from the main body of the dSph.  The clump is the only intermediate-age aggregate of stars detected in a dSph galaxy.  Based on its position, shape and blue colour, \citet{col04} propose the clump is a remnant of an interaction between Fornax and some smaller, gas-rich companion; that is, shell structure.  Although such shell structure is relatively common in large E galaxies, it has not previously been seen in dwarf galaxies.  The spectroscopic study by \citet{pont04} found a significant increase in star formation within Fornax approximately 2 Gyr ago, possibly related to this merger event.  Also, the simulations of \citet{knebe04} appear to support the formation of shells during the interaction of a satellite with its host under the assumption of a non-static DM potential.  Assuming these simulations are relevant at the scale of dwarf galaxies, they correspond well with the shells now being discovered in Fornax.

\section{Conclusion}
Substructre in dwarf galaxies is an important astronomical tool.  The Sagittarius system (and possibly UMi) is undergoing distortion due to the Galactic tidal field, and the resulting tidal tail formation can be used to measure the dark matter content of both the satellite and the Galaxy.  Also, this is a modern example of hierarchical merging, and hence represents a test case for theories of large-scale structure formation.  Recent measurements may have started to detect the accretion history of dwarf galaxies; the UMi and Fornax dwarf galaxies both contain separate, dense stellar associations.  Finally, the simulations of large-scale structure formation do not yet make predictions for the {\em internal} structure of dark sub-halos, however observations support the existence and importance of substructure in dwarf galaxies.


\label{lastpage}

\end{document}